\documentclass[useAMS,usenatbib,usegraphicx]{mn2e}
\title[Gravitational lensing by TIS]
      {Gravitational lensing by the truncated isothermal sphere\\
       model for cosmological halos. I. Individual lens properties}
\author[Hugo Martel and Paul R. Shapiro]
       {Hugo Martel\thanks{E-mail: hugo@simplicio.as.utexas.edu} 
        and Paul R. Shapiro\thanks{E-mail: shapiro@astro.as.utexas.edu}\\
Department of Astronomy, University of Texas, Austin, TX 78712}

\begin{document}

%\date{Accepted 1988 December 15. Received 1988 December 14; 
%in original form 1988 October 11}
\date{Submitted July 29 2002}

\pagerange{\pageref{firstpage}--\pageref{lastpage}} \pubyear{2002}

\maketitle

\label{firstpage}

\begin{abstract}
The gravitational lensing properties of cosmological halos depend
upon the mass distribution within each halo. The description
of halos as nonsingular, truncated isothermal spheres, a 
particular solution of
the isothermal Lane-Emden equation (suitably modified for $\Lambda\neq0$),
has proved to be a useful approximation for the halos which form from
realistic initial conditions in a CDM universe. We derive here the basic
lensing properties of such halos, including the image separation,
magnification, shear, and time-delay. We also provide analytical
expressions for the critical curves and caustics.
We show how the scale-free results we derive yield scale-dependent lensing properties which depend upon the cosmological background universe and
the mass and collapse redshift of the lensing halos, according to
the truncated isothermal sphere (TIS) model of CDM halos derived
elsewhere. We briefly describe the application of these results to
the currently-favored $\Lambda$CDM universe.
\end{abstract}

\begin{keywords}
cosmology: theory -- dark matter --
galaxies: clusters: general -- galaxies: formation -- 
galaxies: halos -- gravitational lensing
\end{keywords}

\section{Introduction}
The gravitational lensing of distant sources has in recent years become
one of the most powerful tools in observational cosmology (see, for example,
\citealt{soucail01} and references therein). Since the
effects of gravitational lensing depend upon the redshift of the source,
the cosmological background, and the distribution of matter in the universe,
they can be used to constrain the cosmological parameters and the 
primordial power spectrum of density fluctuations from which structure
originates. In addition, many of the effects produced by gravitational lenses,
such as image multiplicity, separations, and time delay,
depend strongly upon the matter distribution inside the lenses.
Hence, measurements of these effects can
provide a unique tool for probing the matter
distribution inside collapsed objects like galaxies and clusters,
providing the only direct measurement of their dark matter content, 
and constraining the theory of their 
formation and evolution.

Until recently, the internal structure of halos adopted in lensing
studies was generally some gravitational equilibrium 
distribution, either singular or
nonsingular (e.g., King model, singular isothermal sphere, 
pseudo-isothermal sphere), 
not necessarily motivated directly by the
theory of cosmological halo formation 
(see, e.g., \citealt{young80}; \citealt{tog84}; 
\citealt{hk87}; \citealt{nw88}; \citealt{bsbv91};
\citealt{jaros91,jaros92}; \citealt{kochanek95}; \citealt{pmm98};
\citealt{premadi01}; \citealt{rm01}). As the theory
of halo formation in the CDM model has advanced in recent years, however,
the halo mass profiles adopted for lensing models have been refined to
reflect this theory.
Numerical simulations of large-scale
structure formation in Cold Dark Matter (CDM) universes predict that
galaxies and clusters have a singular density profile
which approaches a power law $\rho\propto r^{-n}$ at the center,
with the exponent $n$ ranging from 1 to 1.5 (\citealt{cl96};
\citealt{nfw96,nfw97}; \citealt{tbw97};
\citealt{fm97,fm01a,fm01b}; \citealt{moore98,moore99}; 
\citealt{hjs99}; \citealt{ghigna00}; \citealt{js00};
\citealt{klypin00}; \citealt{power02}).
These results are in apparent conflict with observations
of rotation curves of dark-matter-dominated dwarf 
galaxies and low surface brightness galaxies,
which favor a flat-density core (cf. \citealt{primack99}; \citealt{bs99}; 
\citealt{moore99}; \citealt{moore01}).
On the scale of clusters of galaxies, observations of 
strong gravitational lensing
of background galaxies by foreground clusters also favor the presence of a
finite-density core in the centers of clusters (see, e.g., \citealt{tkda98}). 

Several possible explanations have been suggested in order
to explain this discrepancy. 
The rotation curve data might lack sufficient spatial resolution 
near the center to distinguish unambiguously between a density profile 
with a flat-density core and one with a singular profile
(e.g. \citealt{vdbs01}).
Attempts have also been made to improve the numerical resolving 
power of the simulations to obtain a more accurate determination
of the slope of the predicted density profiles at small radii
(e.g. \citealt{moore99}; \citealt{power02}).
However, if the flat-core interpretation of the
observations and the singular cusps predicted by the numerical simulations are
both correct, then the simulation algorithms may be ignoring
some physical process which would, if included, serve to
flatten the halo density profiles at small 
radii relative to the results for purely gravitational, N-body
dynamics of cold, collisionless dark matter,
while retaining the more successful aspects of the
CDM model. For example, gasdynamical processes 
(see, e.g. \citealt{esh01}) and a modification
of the microscopic properties of CDM, such as the proposal of
self-interacting dark matter \citep{ss00}, 
both have the potential to lower the central
density of halos and possibly reconcile simulations with observations.

Lensing by the two kinds of halo mass profiles, singular versus flat-core,
will be different. This has led to attempts to predict the
differences expected if the halos have the
singular cusp of the NFW or Moore profiles or else
a profile with a flat core (e.g. \citealt{kochanek95};
\citealt{km01}; \citealt{rm01}; \citealt{wts01};
\citealt{tc01}; \citealt{lo02}). 
Singular profiles like that of NFW are physically motivated by the N-body
simulations, and the latter have been used to place these halo
profiles empirically in a proper
cosmological context which permits statistical predictions for the
CDM model. The nonsingular profiles which have been adopted to contrast
with these singular ones, however,
are generally no more than parameterized, mathematical fitting formulae, with
no particular physical model to motivate them or put them in a proper
cosmological context.

We have developed an analytical model for the postcollapse equilibrium
structure of virialized objects that condense out of a cosmological background
universe, either matter-dominated or flat with a cosmological constant
(\citealt{sir99}, hereafter Paper~I;
\citealt{is01a}, hereafter Paper~II). This
{\it Truncated Isothermal Sphere\/}, or TIS, model assumes
that cosmological halos form from the collapse and virialization of
``top-hat'' density perturbations and are spherical, isotropic, and 
isothermal. This leads to a unique, nonsingular
 TIS, a particular solution of the Lane-Emden equation
(suitably modified when $\Lambda\neq0$). 
The size $r_t$ and velocity dispersion $\sigma_V$ are unique functions of the
mass $M$ and formation redshift $z_{\rm coll}$ of the object for
a given background universe. The TIS density profile flattens to a 
constant central value, $\rho_0$, which is roughly proportional to the
critical density of the universe at the epoch of collapse,
with a small core radius $r_0\approx r_t/30$
(where $\sigma_V^2=4\pi G\rho_0r_0^2$ and $r_0\equiv r_{\rm King}/3$,
for the ``King radius'' $r_{\rm King}$, defined by \citealt{bt87},
p. 228).

Even though the TIS model does not produce the central cusp in the
density profile of halos predicted by numerical CDM simulations at
very small radii,
it does reproduces many of the average properties of these halos 
quite well, suggesting that it is a useful 
approximation for the halos which result from more realistic initial 
conditions (Papers I, II; \citealt{is01b} and
references therein). In particular,
the TIS mass profile agrees well with the fit by NFW to
N-body simulations (i.e. fractional
deviation of $\sim20\%$ or less) at all radii outside of a few TIS core radii
(i.e. outside a King radius or so).
It also predicts the internal structure of X-ray
clusters found by N-body and gasdynamical simulations of cluster 
formation in the CDM model. For example, 
the TIS model reproduces to great accuracy the
mass-temperature and radius-temperature virial relations and integrated 
mass profiles derived empirically from the simulations of cluster formation
\citep{emn96}.
The TIS model also successfully reproduces to high precision the mass-velocity 
dispersion relation for clusters in CDM simulations of
the Hubble volume by the Virgo Consortium \citep{evrard02}, including 
its dependence on redshift for different background cosmologies. The
TIS model also 
correctly predicts the average value of the virial ratio
in N-body simulations of halo formation in CDM.   

The TIS profile matches the observed mass profiles of 
dark-matter-dominated dwarf galaxies. 
The observed rotation curves of
dwarf galaxies are generally well fit by a density profile
with a finite density core suggested by \citet{burkert95}, given by
\begin{equation}
\rho(r)=\frac
{\rho_{0,B}}{(r/r_c+1)(r^2/r_c^2+1)}\,.
\end{equation}

\noindent
The TIS model gives a nearly perfect fit to this profile,
with best fit parameters
$\rho_{0,B}/\rho_{0,{\rm TIS}}=1.216$, $r_{c}/r_{0,{\rm TIS}}=3.134$,
correctly predicting the maximum 
rotation velocity $v_{\rm max}$ and the radius $r_{\rm max}$
at which it occurs.
The TIS  model can also
explain the mass profile with a flat density core 
measured by \citet{tkda98} for cluster CL 0024+1654
at $z=0.39$, using the strong gravitational lensing of background 
galaxies by the cluster to infer the cluster mass distribution
\citep{si01}.
The TIS model not only provides a good fit to the projected 
surface mass density distribution of this cluster within the arcs, but
also predicts the overall 
mass, and a cluster velocity dispersion in close agreement with the value 
$\sigma_v=1150$ km/s measured by \citet{dressler99}.

Several authors have studied the effect of lensing by halos with
a flat-density core (\citealt{jaros91,jaros92}; \citealt{kochanek95};
\citealt{pmm98,premadi01}) or by NFW or Moore profiles
that have been generalized, so
that the inner slope of the density profile is arbitrary
(\citealt{km01}; \citealt{rm01}; \citealt{wts01}; \citealt{lo02}).
These particular density profiles are essentially mathematical
conveniences without physical motivation. There is no underlying
theoretical model in these cases that was used to predict the value of the
core radius or the departure of the inner slope of the density profile
from the value found by N-body simulations of CDM.
By contrast, the TIS model is based on a set of physical
assumptions concerning the origin, evolution, and
equilibrium structure of halos in CDM universes. Observations of gravitational
lenses have the potential to distinguish between the TIS profile
and singular ones like the NFW profile, as several observable properties of 
gravitational lenses will be strongly affected by the presence, or
absence of a central cusp in the density profile. 
One example of an
important observable that can distinguish between various density profiles
is the parity of the number of images. Lenses with nonsingular density
profiles, such as the TIS, obey the {\it odd number theorem}. The number of
images of a given source is always odd, unless the source is extended and
saddles a caustic (see \citealt{sef92}, hereafter SEF,
p.~172). Lenses with singular profiles, like the singular isothermal sphere,
the NFW profile, or the Moore profile, need not obey this theorem, even for
point sources. Most observed multiple-image gravitational lenses
have either 2 or 4 images, and this may argue against 
profiles with a central core \citep{rm01}.
There are, however, other possible explanations for the 
absence of a third or fifth
image. That image tends to be very close to the optical axis, 
and might be hidden behind the lens itself. Also, it 
is usually highly demagnified, and might be too faint to be seen. 

We can use the TIS solution to model observed gravitational
lenses individually. Alternatively, we can use the
observations collectively to constrain the distribution of halo properties
as characterized by the TIS solution. These properties, core radius, velocity 
dispersion, central density, and so on, depend upon the mass of 
the lensing halos and
the redshift at which they form. Observational constraints on
the statistical distribution of these properties will, in turn, impose 
constrains on the cosmological parameters and the 
primordial power spectrum of density fluctuations.

The problem of studying gravitational lensing of distant
sources in an inhomogeneous universe can be divided into two parts. The
first part consists of determining the intrinsic properties of the
lenses. In particular, we need to determine the relationship between
the observables (image multiplicity, magnification, brightness ratio,
sheer, image separation, time delay, $\ldots$) and the lens parameters.
The second part consists of tying the lens properties to the
cosmology. This involves using cosmological models
of structure formation and evolution to determine the statistical
distribution of the lens parameters, the clustering properties of the
lenses, and the nature of
their environments. In this paper, we focus on the first part,
determining the intrinsic properties of individual lenses,
which is an essential building block. The second part will be the
subject of forthcoming papers.

The remainder of this paper is organized as follows. In \S2, we derive
the lens equation. In \S3, we compute the critical curves and caustics.
In \S4, we study the properties of multiple images: separation,
magnification, brightness ratios, and time delay. In \S5, we place the
scale-free description of these properties in the cosmological
context of the CDM model and explain how the dimensionless
parameters of \S2--\S4 are related by the TIS model to the properties
of lensing halos in physical units.
Summary and conclusion are presented in \S6.

\section{THE LENS EQUATION}

Figure 1 illustrates the lensing geometry. $\eta$ and $\xi$ are the position
of the source on the source plane and the image on the image plane, respectively,
$\hat\alpha$ is the deflection angle, and $D_L$, $D_S$, and $D_{LS}$ are
the angular diameter distances between observer and lens, observer and source,
and lens and source, respectively. The lens equation is
\begin{equation}
\eta={D_S\over D_L}\xi-D_{LS}\hat\alpha\,.
\end{equation}

\begin{figure}
\vspace{-40pt}
\hspace{-45pt}
\includegraphics[width=106mm]{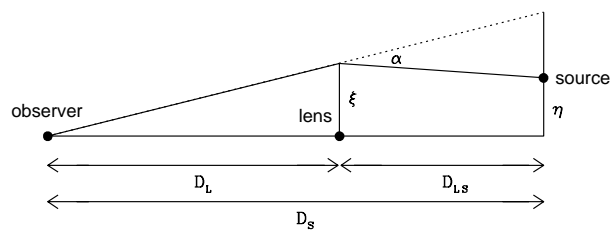}
 \vspace{-160pt}
 \caption{The lensing geometry: the dots indicate the location of the 
observer, lensing galaxy, and source. $r_c$ is the core radius of the galaxy, 
and $\eta$ is the distance between the source and the optical axis.
The angular diameter distances $D_L$, $D_{LS}$, and $D_S$ are also indicated.}
\end{figure}

\noindent Notice that since the lens is axially symmetric, we can write the
quantities $\eta$, $\xi$, and $\hat\alpha$ as scalars instead
of 2-component vectors.
We introduce a characteristic length scale $\xi_0$, and rescale the
positions and deflection angle, as follows:
\begin{eqnarray}
y\!\!\!\!&=&\!\!\!\!{D_L\eta\over D_S\xi_0}\,,\\
\label{xscale}
x\!\!\!\!&=&\!\!\!\!{\xi\over\xi_0}\,,\\
\alpha\!\!\!\!&=&\!\!\!\!{D_LD_{LS}\hat\alpha\over D_S\xi_0}\,.
\end{eqnarray}

\noindent The lens equation reduces to
\begin{equation}
y=x-\alpha(x)\,.
\end{equation}

\noindent To compute the deflection angle $\alpha$, we first need an 
expression for
the projected surface density $\sigma(\xi)$.
The density profile of the TIS is well-fitted by the following approximation:
\begin{equation}
\label{tisfit}
\rho(r)=\rho_0\left({A\over a^2+r^2/r_0^2}-{B\over b^2+r^2/r_0^2}\right)
\end{equation}

\noindent (Paper~I) 
where $A=21.38$, $B=19.81$, $a=3.01$, $b=3.82$. The projected
surface density is given by
\begin{equation}
\label{proj}
\sigma(\xi)=\int_{-\infty}^\infty\rho(r)dz\,,
\end{equation}

\noindent where $\xi$ is the projected distance from the center of the TIS, 
and $z=(r^2-\xi^2)^{1/2}$. This expression assumes that the fit given 
by equation~(\ref{tisfit}) 
is valid all the way to $r=\infty$. Actually, it is only valid up to a
truncation radius $r_t\approx30r_0$ (the actual
value is 29.4 for an Einstein-de~Sitter universe [Paper I],
slightly different for an open matter-dominated universe, or a
flat universe with a cosmological constant [Paper~II]).
One could always set the limits 
in equation~(\ref{proj})
to $\pm(r_t^2-\xi^2)^{1/2}$. However, it turns out that
the change in $\sigma$ would be small,
simply because the value of 30 is significantly larger than both $a$ and $b$.
Also, the density $\rho(r)$ is nonzero at $r>30r_0$, 
and ignoring its contribution entirely
would be incorrect. For the sake of simplicity, we shall assume that 
equation~(\ref{proj}) 
remains a good approximation out to $r=\infty$. We substitute 
equation~(\ref{tisfit}) in equation~(\ref{proj}), and get
\begin{equation}
\label{sigma}
\sigma(\xi)=\pi\rho_0r_0^2\left[{A\over(a^2r_0^2+\xi^2)^{1/2}}-
{B\over(b^2r_0^2+\xi^2)^{1/2}}\right]
\end{equation}

\noindent [This result was also derived by \citet{nlb97}
and \citet{iliev00}].
For spherically symmetric lenses, the deflection angle is given by
\begin{equation}
\label{alpha1}
\alpha(x)={2\over x}\int_0^xx'{\sigma(x')\over\sigma_{\rm crit}}dx'
\end{equation}

\noindent [SEF, eq.~(8.3)], where $\sigma_{\rm crit}$ is
the critical surface density, given by
\begin{equation}
\label{sigmacrit}
\sigma_{\rm crit}={c^2D_S\over4\pi GD_LD_{LS}}\,,
\end{equation}

\noindent where $c$ and $G$ are the speed of light and the gravitational constant,
respectively, and $D_L$, $D_S$, and $D_{LS}$ are
the angular diameter distances between observer and lens,
observer and source, and lens and source, respectively. 
We substitute equation~(\ref{sigma}) into equation~(\ref{alpha1}),
eliminate~$\xi$ using equation~(\ref{xscale}), and set the 
characteristic scale $\xi_0$ equal to $r_0$. We get
\begin{eqnarray}
\alpha(x)\!\!\!\!&=&\!\!\!\!{2\pi\rho_0r_0\over\sigma_{\rm crit}x}
\big[A(a^2+x^2)^{1/2} \nonumber \\
\label{alpha2}
&&\qquad\qquad\qquad-\,B(b^2+x^2)^{1/2}-Aa+Bb\big]\,.
\end{eqnarray}

\noindent 
We now introduce the
dimensionless central surface density, or central convergence, 
$\kappa_c$, defined by
\begin{equation}
\kappa_c\equiv{\sigma(\xi=0)\over\sigma_{\rm crit}}
={\pi\rho_0r_0\over\sigma_{\rm crit}}\left({A\over a}-{B\over b}\right)\,,
\end{equation}

\noindent and use this expression to eliminate $\sigma_{\rm crit}$ in
equation~(\ref{alpha2}). It reduces to
\begin{eqnarray}
\alpha(x)\!\!\!\!&=&\!\!\!\!{2ab\kappa_c\over(Ab-Ba)x}
\big[A(a^2+x^2)^{1/2} \nonumber \\
\label{alpha3}
&&\qquad\qquad\qquad-\,B(b^2+x^2)^{1/2}-Aa+Bb\big]\,.
\end{eqnarray}

\noindent This expression has the following limiting cases:
\begin{equation}
\label{alpha4}
\alpha(x)=\cases{
\kappa_cx\,,&$x\ll a$, $b$\,;\cr
\displaystyle{2ab(A-B)\kappa_c\over Ab-Ba}\,,&$x\gg a$, $b$\,.\cr}
\end{equation}

\noindent The final form of the lens equation is
\begin{eqnarray}
y\!\!\!\!&=&\!\!\!\!x-{2ab\kappa_c\over(Ab-Ba)x}
\big[A(a^2+x^2)^{1/2} \nonumber \\
\label{lensfinal}
&&\qquad\qquad\qquad-\,B(b^2+x^2)^{1/2}-Aa+Bb\big]\,.
\end{eqnarray}

\noindent
[This result was also obtained by \citet{ct01}].

\section{CRITICAL CURVES AND CAUSTICS}

\subsection{Solutions}

The determination of the critical curves is quite trivial for
axially symmetric lenses.
The dimensionless interior mass $m(x)$ is related to the deflection angle
$\alpha(x)$ by
\begin{eqnarray}
m(x)\!\!\!\!&\equiv&\!\!\!\!\alpha(x)x={2ab\kappa_c\over(Ab-Ba)}
\big[A(a^2+x^2)^{1/2} \nonumber \\
\label{mx}
&&\qquad\qquad\qquad-\,B(b^2+x^2)^{1/2}-Aa+Bb\big]\,.
\end{eqnarray}

\noindent [SEF, eq.~(8.3)]. Tangential critical curves are defined
by
\begin{eqnarray}
{m(x)\over x^2}\!\!\!\!&=&\!\!\!\!{2ab\kappa_c\over(Ab-Ba)x^2}
\big[A(a^2+x^2)^{1/2} \nonumber \\
\label{tancrit}
&&\qquad\qquad-\,B(b^2+x^2)^{1/2}-Aa+Bb\big]=1\,.
\end{eqnarray}

\noindent This can be turned into a 
fourth-degree equation for $x^2$. Even though such equation can
be solved analytically, this is a case where a numerical solution is
preferable. But first let us investigate the existence of a solution.
According to equations~(\ref{alpha4}) and~(\ref{mx}), 
$m(x)=\kappa_cx^2$ at small $x$.
As $x$ increases, the dependence of $m(x)$ on $x$ drops from $m(x)\propto x^2$
to $m(x)\propto x$. If $\kappa_c<1$, $m(x)$ ``starts up'' below $x^2$ and
can never raise above it, and therefore equation~(\ref{tancrit}) 
has no solution. If $\kappa_c=1$, equation~(\ref{tancrit}) 
has a unique solution at $x=0$. Finally, if
$\kappa_c>1$, there is a unique solution $x>0$. This is illustrated in
the top panel of
Figure~2. This result was expected: critical curves are associated
with the phenomenon of multiple imaging. For axially symmetric
lenses with monotonically decreasing surface density
($d\sigma/d\xi<0$), multiple images can occur only if
$\kappa_c>1$ (SEF, p.~238). 

\begin{figure}
\vspace{-10pt}
\hspace{-20pt}
\includegraphics[width=100mm]{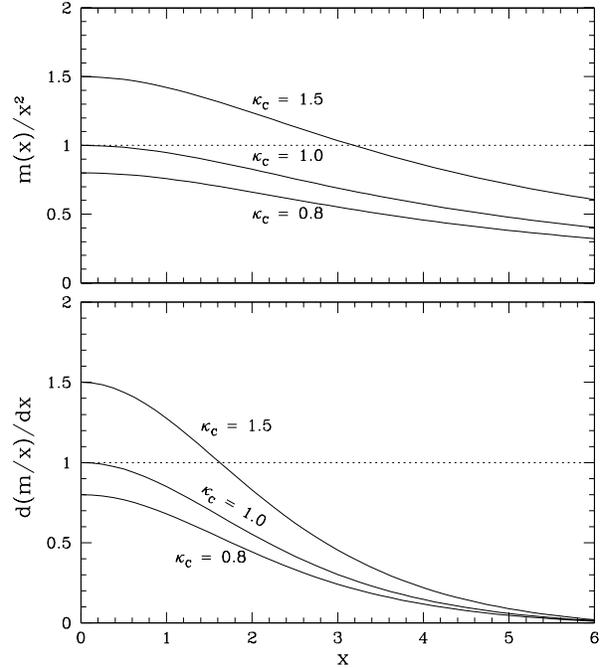}
 \vspace{-15pt}
 \caption{Top: $m(x)/x^2$ versus $x$, for 3 particular values of
$\kappa_c$. Tangential critical curves, defined by $m(x)/x^2=1$, can only
occur for $\kappa_c>1$.
Bottom: $d(m/x)/dx$ versus $x$, for 3 particular values of
$\kappa_c$. Radial critical curves, defined by $d(m/x)/dx=1$, can only
occur for $\kappa_c>1$.}
\end{figure}

Radial critical curves are defined by
\begin{eqnarray}
{d(m/x)\over dx}\!\!\!\!&=&\!\!\!\!{2ab\kappa_c\over(Ab-Ba)x^2}
\bigg[-{Aa^2\over(a^2+x^2)^{1/2}} \nonumber \\
\label{radcrit}
&&\qquad+\,{Bb^2\over(b^2+x^2)^{1/2}}+Aa-Bb\bigg]=1\,.
\end{eqnarray} 

According to equation~(\ref{alpha4}), 
$d(m/x)/dx=d\alpha/dx=\kappa_c$ at small $x$.
Equation~(\ref{radcrit}) 
clearly shows that $d(m/x)/dx$ is a monotonically decreasing
function of $x$. Hence, equation~(\ref{radcrit}) 
has a nonzero solution only if
$\kappa_c>1$. This is illustrated in the bottom panel of Figure~2.

We have solved equations~(\ref{tancrit}) and~(\ref{radcrit}) numerically 
for the tangential critical radius $x_t$ and radial critical radius~$x_r$.
The solutions are plotted
in Figure~3, as functions of $\kappa_c$. Also plotted is the radial caustic
radius $y_r$, obtained by substituting the value of $x_r$ into 
equation~(\ref{lensfinal}).
(The value of $y_r$ we obtain is actually negative, but 
this reflects the fact that the equation locates the point in the
plane of the sky which is the intersection between the radial caustic
circle projected onto that plane and a line in this plane from some point on
the radial critical circle through the origin at the center of the lens;
the intersection point on the radial caustic circle is on the
opposite side of the origin. The actual radius of the caustic circle,
then, is the absolute value of $y_r$ in that case.)
Both $x_t$ and $y_r$ increase rapidly with $\kappa_c$, while the
value of $x_r$ levels off. We can easily find the asymptotic behavior in the
limit of large $\kappa_c$. To find $x_t$, we take the limit 
$\kappa_c\rightarrow\infty$ in equation~(\ref{tancrit}). 
To satisfy this equation,
either the factor $x^2$ in the denominator must diverge, or the quantity
in bracket must vanish. But this quantity is zero only for $x=0$, and
approaches 0 as $x^2$, cancelling the $x^2$ in the denominator. Hence, the
only way to satisfy equation~(\ref{tancrit}) 
in the limit $\kappa_c\rightarrow\infty$
is by having $x\rightarrow\infty$ in the denominator. 
We take the limit $x\rightarrow\infty$ in equation~(\ref{tancrit}), and
keep the first and second leading terms. We get
\begin{equation}
\label{xt1}
x={2ab\kappa_c\over Ab-Ba}\left(A-B+{Bb-Aa\over x}\right)\,.
\end{equation}

\begin{figure}
\vspace{-50pt}
\hspace{-30pt}
\includegraphics[width=106mm]{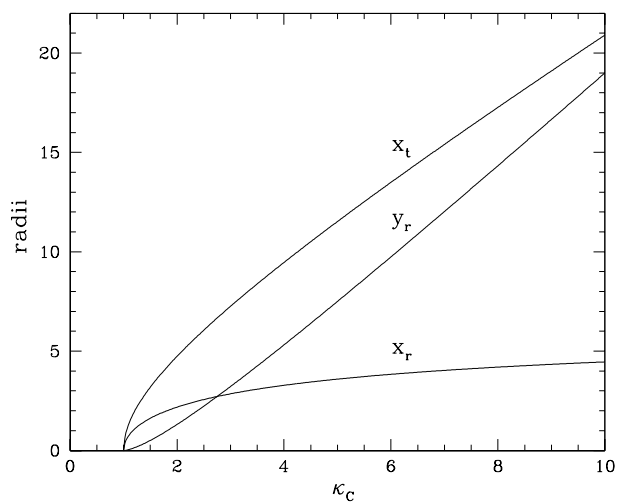}
 \vspace{-50pt}
 \caption{Radii of the radial critical circle, $x_r$, tangential
critical circle, $x_t$, and radial caustic, $y_r$, versus $\kappa_c$.}
\end{figure}

\noindent Setting $x=\infty$ in the last term, we get
$x\approx2ab(A-B)\kappa_c/(Ab-Ba)$. We then substitute this
value in the last term of equation~(\ref{xt1}), and get
\begin{eqnarray}
x_t\!\!\!\!&=&\!\!\!\!{2ab(A-B)\over Ab-Ba}\kappa_c+{Bb-Aa\over A-B}
 \nonumber \\
\label{xt2}
&&\qquad\qquad=1.638\kappa_c+7.210\,,\qquad\kappa_c\gg1\,.
\end{eqnarray}

\noindent To find $x_r$, we take the limit $\kappa_c\rightarrow\infty$ in
equation~(\ref{radcrit}). 
In this case, the quantity in brackets first increases with
$x$, then decreases and drops to 0 at a finite value $x=6.479$. Hence,
unlike equation~(\ref{tancrit}), it is the bracket, not the factor of $x^2$ in
the denominator, that prevents the expression from diverging as
$\kappa_c\rightarrow\infty$. The asymptotic limit is therefore
\begin{equation}
\label{xr}
x_r=6.479\,,\qquad\kappa_c\gg1\,.
\end{equation}

\noindent Finally, we substitute this value of $x_r$ in 
equation~(\ref{lensfinal}),
and take the limit $\kappa\rightarrow\infty$. We get
\begin{equation}
y_r=2.425\kappa_c-6.479\,,\qquad\kappa_c\gg1\,.
\end{equation}

\subsection{Illustrative example}

Using a simple ray-tracing algorithm, we computed the image(s) of a circular
source of diameter $\Delta y=1$, created by a TIS with central convergence
$\kappa_c=4.015$. The results are shown in Figure~4 for 8 different
locations of the source, ranging from $y=8.0$ to $y=0.0$. For each case, the
left panel shows the source and the caustic circle ($y_r=5.640$) on
the source plane, and the right panel shows the images(s), the radial
critical circle ($x_r=3.334$), and the tangential critical circle
($x_t=9.783$) on the image plane. For the cases $y=8.0$ and $6.0$,
only one image appears. At $y=5.4$, the source overlaps the caustic, and a
second, radially-oriented image appears on the radial critical circle.
At $y=4.8$, the source is entirely inside the caustic, and the second
image splits in two images, located on opposite sides of the radial
critical circle, forming with the original image a system of 3 aligned images.
As the source moves toward $y=0$, the central image moves toward $x=0$
and becomes significantly fainter, while the other images move toward the
tangential critical circle and become bright, elongated arcs. At $y=0$,
the two arcs have merged to form an Einstein ring located on top of
the tangential critical circle, while the central image, very faint,
is still visible in the center.

\begin{figure*}
\vspace{-30pt}
\hspace{-40pt}
\includegraphics[width=190mm]{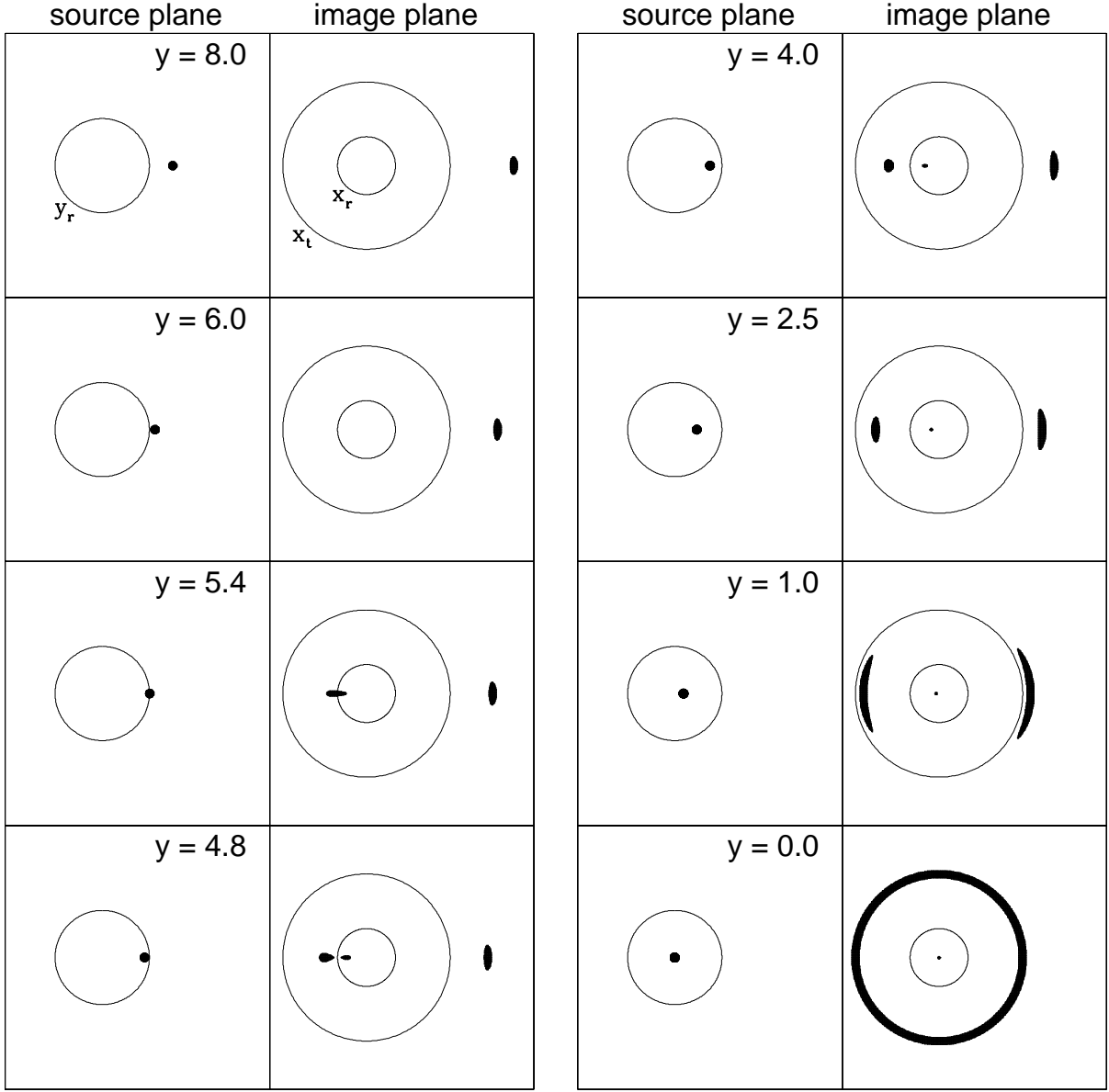}
\vspace{-55pt}
 \caption{Images of a circular source. Each pair of panels shows
the source plane in the left panel, with the caustic, and the
image plane in the right panel, with the radial (inner) and
tangential (outer) critical circles. The position $y$ of the source on
the source plane is indicated. We used $\kappa_c=4.014$, and a source
of diameter $\Delta y=1$.}
\end{figure*}

The central image is always located inside the radial critical curve,
which has a maximum radius of 6.479 
(eq.~[\ref{xr}]). The density profile of the 
TIS extends to a cutoff radius of order 29.4 (Papers~I and II)
[Note: We are neglecting here the effect of matter distributed outside
the virial radius of our lensing halo.]. 
Hence the central image is
always located ``inside'' the TIS and will be visible only if the lens
is transparent. Incidently, all the images shown in Figure~4 are located
inside the TIS, because the value we used for $\kappa_c$ is rather modest.
Setting $x_t=29.4$ in equation~(\ref{xt2}), 
we get $\kappa_c=17.9$. This is the value
of $\kappa_c$ for which the tangential critical circle is located along the
edge of the TIS. For opaque lenses, the image located between the tangential
and radial critical circles (leftmost 
images in Fig.~4) might be visible only
if $\kappa_c>17.9$. For $\kappa_c<17.9$, only the outermost
image, located outside the tangential critical circle, might be visible.

\section{PROPERTIES OF MULTIPLE IMAGES}

\subsection{Image separation}

The locations of the images are computed by solving the lens 
equation~(\ref{lensfinal}).
After rewriting this equation as $x-y=\alpha(x)$, we can solve it graphically.
In Figure~5, we show the multiple image diagram for the particular case
$\kappa_c=5.005$. The solid curves shows $\alpha(x)$, while the
dotted lines show $x-y$ for particular values of $y$. Each intersection
between a line and the curve corresponds to one image. If $y>y_r$ (bottom
line), the source is outside the caustic circle, and only one image appears.
If $y=y_r$ the source is on the caustic circle
and a second image appears on the radial
critical circle, at $x=-x_r$. For $y<y_r$, the source is inside the caustic,
and the second image splits into two images. Finally, for $y=0$ (top curve),
the central image is located at $x=0$, and the two outer images are located 
on the tangential circle, at $x=\pm x_t$. Actually, these two images merge
to form an Einstein ring. The slope of the curve $\alpha(x)$ versus $x$
is equal to $\kappa_c$ at $x=0$. It is clear from Figure~5 that if we
lower $\kappa_c$ below 1, any $x-y$ versus $x$ line will intersect the
curve only once; multiple images cannot occur if $\kappa_c<1$. In Figure~6,
we plotted the solution of the lens equation for various values of $\kappa_c$.
The bifurcation from 1 to 3 images is clearly visible on all panels with
$\kappa_c>1$. 

\begin{figure}
\vspace{-30pt}
\includegraphics[width=90mm]{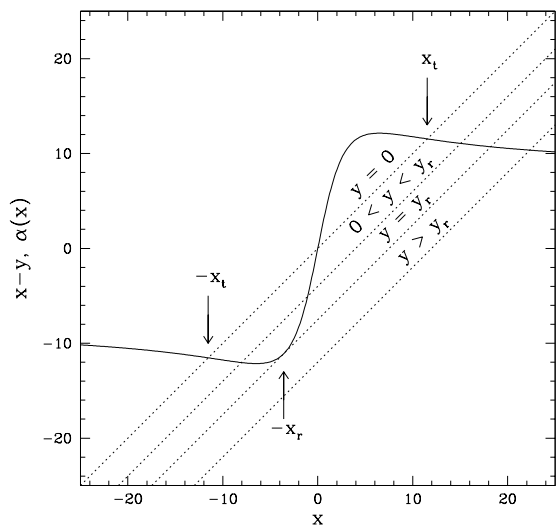}
 \vspace{-20pt}
 \caption{Plot of $\alpha(x)$ (solid curve) and $x-y$ (dotted lines)
versus $x$, for a TIS with $\kappa_c=5.005$ and 4 particular values
of $y$: $y=0$ (top line), $y=4$,
$y=y_r=7.515$, and $y=12$ (bottom line). 
Images are located at values of $x$ corresponding to
intersections between the lines and the curve. Particular solutions,
corresponding to images located on critical curves, are indicated by arrows.}
\end{figure}

\begin{figure}
\hspace{-30pt}
\includegraphics[width=120mm]{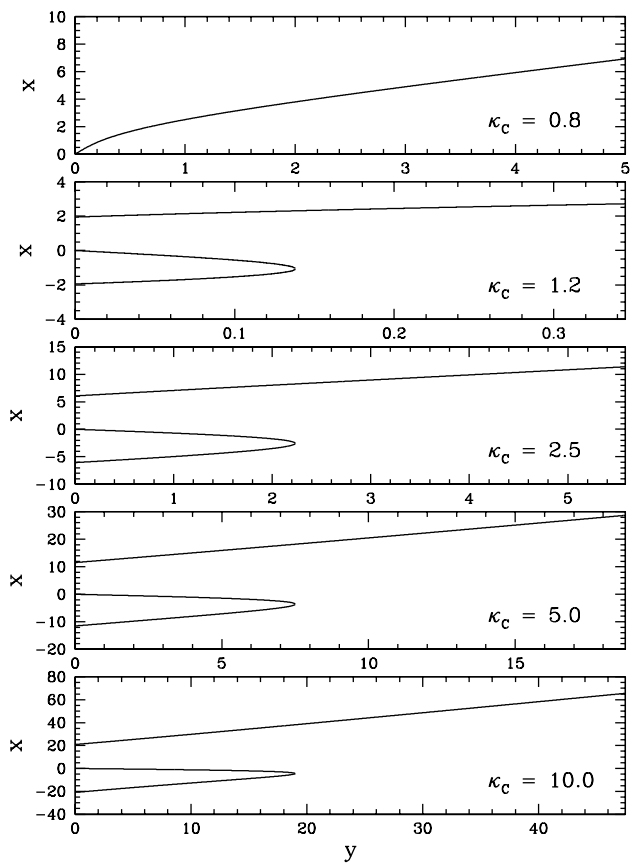}
 \caption{Location $x$ of the image(s) on the image plane versus
source location $y$ on the source plane, for various values of
$\kappa_c$.}
\end{figure}

In Figure~7, we plot the separation between the two outer images as
a function of the source location, for
various values of $\kappa_c$. The plot only extends to $y/y_r=1$, since
larger values of $y$ only produce one image. The separation is fairly
insensitive to the source location, and stays within $\sim15\%$ of the
Einstein ring diameter $\Delta x=2x_t$ for all values of $\kappa_c$
considered. This is particularly convenient for theoretical studies, when
the actual source location can be ignored. 

\begin{figure}
\vspace{-61pt}
\includegraphics[width=90mm]{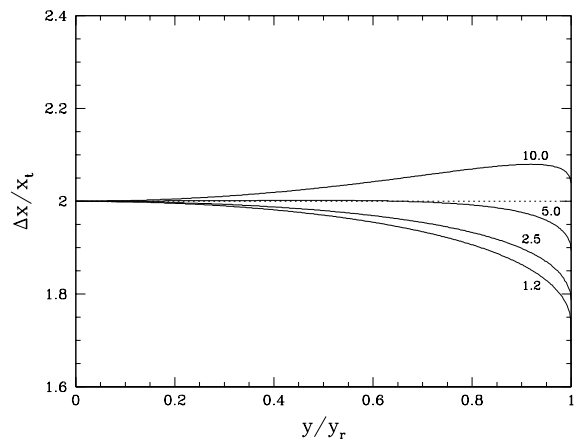}
 \vspace{-30pt}
 \caption{Separation $\Delta x$ between the two outer images,
in units of the tangential critical radius $x_t$, versus source
location $y$ in units of the caustic radius $y_r$. The solid curves
correspond to various values of $\kappa_c$, as indicated. The dotted
line at $\Delta x=2x_t$ indicates the diameter of the Einstein ring.}
\end{figure}

\subsection{Magnification and brightness ratios}

The magnification of an image located at position $x$ on the lens
plane is given by
\begin{eqnarray}
\mu\!\!\!\!&=&\!\!\!\!
\left(1-{m\over x^2}\right)^{-1}
\left[1-{d\over dx}\left({m\over x}\right)\right]^{-1} \nonumber \\
\label{magni}
&&\qquad\qquad\qquad=\left(1-{\alpha\over x}\right)^{-1}
\left(1-{d\alpha\over dx}\right)^{-1}
\end{eqnarray}

\begin{figure}
\hspace{-20pt}
\includegraphics[width=100mm]{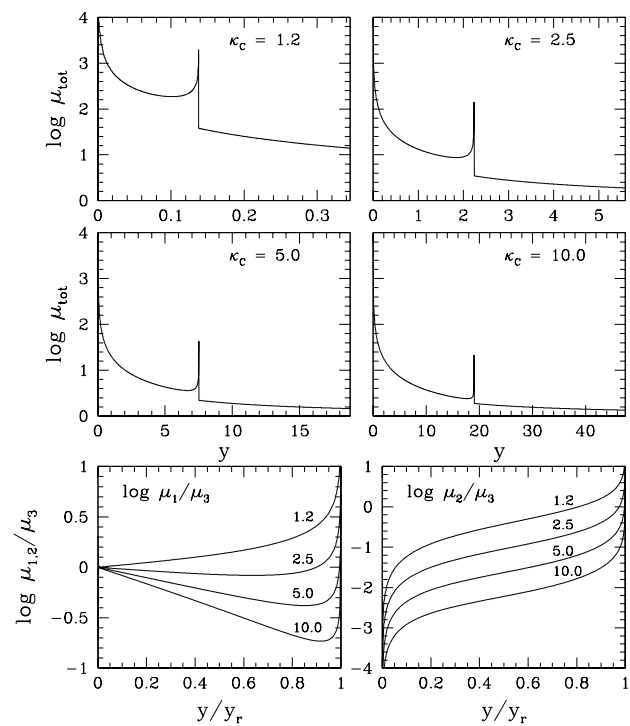}
 \vspace{-20pt}
 \caption{Top four panels: Total magnification $\mu_{\rm tot}$ versus
source position $y$, for four different values of $\kappa_c$.
Bottom panels: Brightness ratios $\mu_1/\mu_3$ and $\mu_2/\mu_3$
versus source position in units of caustic radius, $y/y_r$. The
number next to each curve gives the value of $\kappa_c$.}
\end{figure}

\noindent (SEF, eq.~[8.17]).
We computed the total magnification and brightness ratios of images,
for the four particular cases $\kappa_c=1.2$, 2.5, 5.0, and 10.0. The
results are shown in Figure~8. The top four panels show the total magnification
as a function of the source location $y$. As $y$ decreases, the magnification
slowly increases, until the sources reaches the radial caustic $y=y_r$.
At that moment, a second image, with infinite magnification 
($1-d(m/x)/dx=0$ in eq.~[\ref{magni}]) 
appears on the radial critical curve. As $y$
keeps decreasing, that second image splits into two images, and the 
total magnification becomes finite again, until the source reaches
$y=0$, and an Einstein ring with infinite magnification
($1-m/x^2=0$ in eq.~[\ref{magni}]) appears on the tangential critical curve.
Of course, these infinite magnifications are not physical, since they can only
occur for point sources.

The total magnification is always larger than unity, and always larger
when 3 images are present. Figure~8 shows that the total magnification
decreases with increasing $\kappa_c$, which seems counter-intuitive.
Notice, however, that the top four panels in Figure~8 are plotted for
different ranges of $y$. At a fixed $y$, the total magnification 
always increases with $\kappa_c$.

The bottom panels of Figure~8 shows the brightness ratios $\mu_1/\mu_3$,
and $\mu_2/\mu_3$, where $\mu_i$ is the magnification
of image~$i$, as a function of source position. By convention, image 1 is
the one between the tangential and radial critical curves, image 2
is the one inside the radial critical curve, and image 3 is the one
outside the tangential critical curve (see Fig.~4; from left to right,
the three images are image 1, 2, and 3). For most values of $\kappa_c$,
image~3 is the brightest, image~1 is fainter, but comparable, 
and image~2 (the central one) is much fainter. This is a generic properties
of most axially symmetric lenses with a central core. However, we find
a different behavior at small values of $\kappa_c$. Not only can image~1
become brighter than image~3, but for a source located at $y\la y_r$,
just inside the caustic, even image~2 can become brighter than image~3.

\subsection{Shear}

The total shear $\gamma(x)$ of an image located at position $x$ is given by
\begin{equation}
\label{shear}
\gamma=\left|{m(x)\over x^2}-\kappa(x)\right|\,,
\end{equation}

\noindent (SEF, eq.~[8.15]), where 
$\kappa(x)\equiv\sigma(x)/\sigma_{\rm crit}$. We substitute 
equations~(\ref{sigma}) and~(\ref{mx}) into equation~(\ref{shear}), and get
\begin{eqnarray}
\gamma\!\!\!\!&=&\!\!\!\!{ab\kappa_c\over Ab-Ba}\bigg|
{2A\over x^2}\left[(a^2+x^2)^{1/2}-a\right]
-{2B\over x^2}\big[(b^2+x^2)^{1/2} \nonumber \\
&&\qquad-\,b\big]
-{A\over(a^2+x^2)^{1/2}}+{B\over(b^2+x^2)^{1/2}}\bigg|\,.
\end{eqnarray}

\noindent [This result was also obtained by \citet{nlb97}.]
Figure~9 shows $\gamma$ versus source position~$y$ for various values of 
$\kappa_c$. Notice that the indices 1, 2, 3 refer to the individual images,
and not the the shear components. The shear increase with $\kappa_c$, 
as expected. At large values of $\kappa_c$, $\gamma_1$ increase and 
$\gamma_3$ decrease as $y$ increase, while at small $\kappa_c$ this
trend is reversed. $\gamma_2$ always increase with $y$. In the limit
$y\rightarrow0$, we get $\gamma_1=\gamma_3$ and $\gamma_2=0$;
Images 1 and 3 become identical, while image 2 becomes circular. 

\subsection{Time delay}

For axially symmetric lenses, the deflection potential $\psi(x)$ is
defined by $\alpha=d\psi/dx$. From equation~(\ref{alpha3}), we get
\begin{eqnarray}
\!\!\psi(x)\!\!\!\!\!\!\!\!\!\!\!\!&&=
{2ab\kappa_c\over Ab-Ba}\Big\{
A(a^2+x^2)^{1/2}-Aa\ln\left[a+(a^2+x^2)^{1/2}\right] \nonumber \\
&&-B(b^2+x^2)^{1/2}+Bb\ln\left[b+(b^2+x^2)^{1/2}\right]\Big\}\,.
\end{eqnarray} 

\begin{figure}
\vspace{-25pt}
\hspace{-20pt}
\includegraphics[width=95mm]{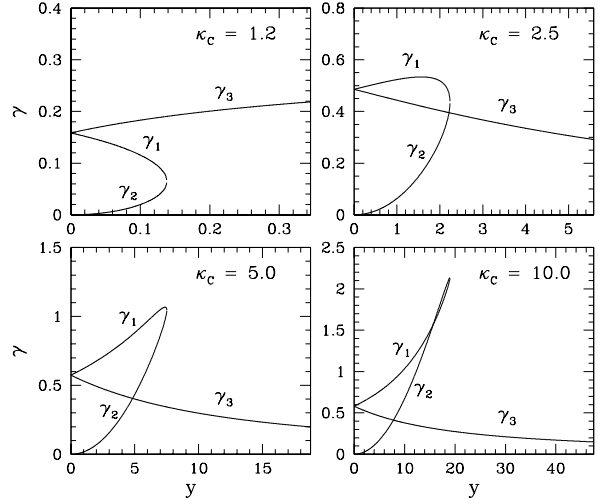}
 \vspace{-50pt}
 \caption{Total shear $\gamma$ versus source position $y$,
for four different values of $\kappa_c$. The indices identify
the various images.}
\end{figure}

\noindent The time delay $\Delta t$ between two images located at $x_i$
and $x_j$, of a source located at $y$, is given by
\begin{equation}
\label{delay}
\Delta t(y)={r_0^2D_S\over cD_LD_{LS}}
(1+z_L)\big[\phi(x_i,y)-\phi(x_j,y)\big]
\end{equation}

\noindent (SEF, eq.~[5.44]), where $z_L$ is the
redshift of the lens, and $\phi(x,y)$ is the Fermat potential,
defined by
\begin{equation}
\phi(x,y)={1\over2}(x-y)^2-\psi(x)\,.
\end{equation}

\noindent In equation~(\ref{delay}), the
quantities $D_S$, $D_L$, $D_{LS}$, and $z_L$ depend
on the redshifts of the source and lens, and the background
cosmological model. Only the quantity in square bracket depends on the
intrinsic properties of our lens model. We define
\begin{equation}
\tau_{ij}\equiv\phi(x_i,y)-\phi(x_j,y)\,.
\end{equation}

\noindent Figure~10 shows the time delays as functions of the source location
for the cases $\kappa_c=1.2$, 2.5, 5.0, and 10.0. In all cases, we have
$\tau_{23}>\tau_{13}\geq0$. Hence,
the light from the three images always reach the observer in the
same order, first image~3, then image~1, and finally image~2.
As $y\rightarrow0$, $\tau_{13}\rightarrow0$ because images~1 and~3
move toward the tangential critical circle.
Both $\tau_{13}$ and $\tau_{23}$ increase with $y$, except for small
values of $\kappa_c$, for which $\tau_{23}$ reaches a maximum at
some value $y<y_r$, and then decreases slightly. At $y=y_r$, we
have $\tau_{13}=\tau_{23}$ because images~1 and~2 merge on the radial
critical curve.

\begin{figure}
\vspace{-20pt}
\hspace{-20pt}
\includegraphics[width=90mm]{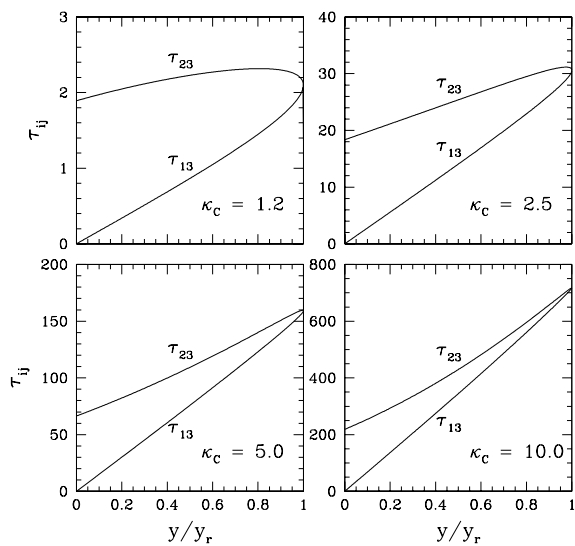}
 \vspace{-30pt}
 \caption{Time delays $\tau_{12}$ and $\tau_{13}$ versus source
position $y$ in units of the caustic radius $y_r$, for four
different values of $\kappa_c$.}
\end{figure}

An interesting quantity is $\tau_{\max}$, the maximum time delay that
a lens can produce. For simplicity, we set $\tau_{\max}\equiv\tau(y=y_r)$,
which, according to Figure~10, is correct for large $\kappa_c$, and
a fairly good approximation at small $\kappa_c$. In Figure~11, we
plot $\tau_{\max}$ as a function of $\kappa_c$. The solution has
the following asymptotic behavior,
\begin{equation}
\label{delaymax}
\tau_{\max}\approx8.3(\kappa_c-1)^2\,,\qquad\kappa_c\gg1\,.
\end{equation}

\begin{figure}
\vspace{-15pt}
\hspace{-10pt}
\includegraphics[width=93mm]{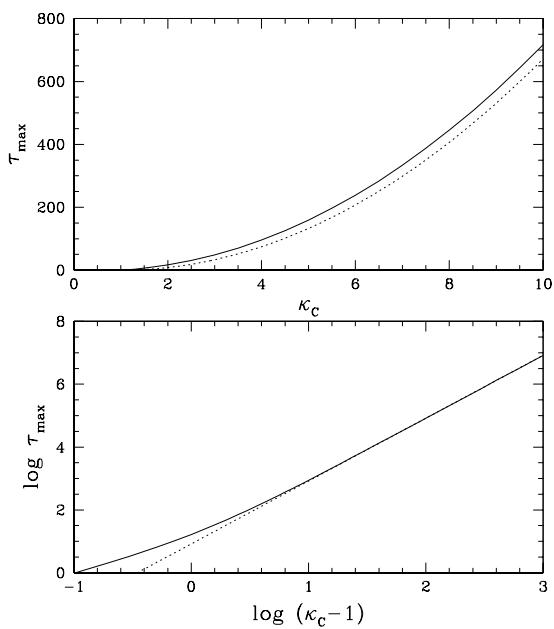}
 \caption{Maximum time delays $\tau_{\max}$ versus $\kappa_c$,
on linear and logarithmic scales (solid curves). The dotted curves
show the fitting formula given by equation~(\ref{delaymax}).}
\end{figure}

\noindent At small $\kappa_c$ (but still larger than unity), 
$\tau_{\max}$ exceeds the value given by
equation~(\ref{delaymax}).

\section{DISCUSSION}

So far, we have focused on the intrinsic, scale-free
properties of the lenses, without
considering the dependence on the cosmological model. All results have
been expressed in terms of two dimensionless parameters, $\kappa_c$ and
$y$. The cosmological models enter the picture when one tries to
determine the typical values and distributions of these parameters.
The value of $\kappa_c$ depends on the critical surface density
$\sigma_{\rm crit}$ (eq.~[\ref{sigmacrit}]), 
which is a function of the lens redshift,
source redshift, and cosmological parameters. In addition,
the TIS profile
parameters $\rho_0$ and $r_0$ are functions of the mass of the object and
its epoch of collapse, which also depend upon the cosmological parameters.
Finally, the distribution of source locations $y$ will be related to
the number densities of sources and lenses, 
which themselves depend upon the cosmological model.

In this section, we will estimate the typical values of $\kappa_c$
for TIS halos in a CDM universe.
We consider the currently-favored $\Lambda$CDM model
with density parameter $\Omega_0=0.3$,
cosmological constant $\lambda_0=0.7$, and Hubble constant 
$H_0=70\,\rm km\,s^{-1}\,Mpc^{-1}$ (or $h=0.7$), and we shall
assume for illustrative purposes
that the source is located at redshift $z_S=3$.
The critical surface density given by equation~(\ref{sigmacrit}) 
depends upon the
lens redshift $z_L$, and has a minimum value at $z_L=0.612$,
given by
\begin{equation}
\label{sigmacritmin}
(\sigma_{\rm crit})_{\min}=0.386\,{\rm g\,cm^{-2}}\,.
\end{equation}

\noindent The surface density of the TIS is computed by plugging
the expressions for $\rho_0$ and $r_0$ provided in Paper~II
into equation~(\ref{sigma}). 
The central value of the surface density is given by
\begin{eqnarray}
\sigma(\xi\!\!\!\!&=&\!\!\!\!0)=\pi\rho_0r_0\left({A\over a}-{B\over b}\right) \nonumber \\
\label{sigma0}
&=&\!\!\!\!0.0400[F(z_{\rm coll})]^2h^{4/3}\left({M\over10^{12}{\rm M_\odot}}
\right)^{1/3}{\rm g\,cm^{-2}}\,.
\end{eqnarray}

\noindent In equation~(\ref{sigma0}), the function $F(z_{\rm coll})$ is a
scaling factor that is used to express the TIS solution for arbitrary
cosmological models in terms of the solution for an Einstein-de~Sitter model.
We will further assume that the lens located at redshift $z_L$ collapsed
at redshift $z_{\rm coll}=z_L$. 

Let us consider the illustrative case of a lens at $z_L=0.612$.
Using the expressions from
Paper~II, we get $F(z_{\rm coll})=1.1644$. 
Equation~(\ref{sigma0}),
with $h=0.7$, reduces to
\begin{equation}
\label{sigma0b}
\sigma(\xi=0)=0.0337\left({M_0\over10^{12}{\rm M_\odot}}\right)^{1/3}
{\rm g\,cm^{-2}}\,.
\end{equation}

\noindent Taking the ratio of equations (\ref{sigma0b}) and 
(\ref{sigmacritmin}), we get
\begin{equation}
\label{kclcdm}
\kappa_c=0.0874\left({M\over10^{12}{\rm M_\odot}}\right)^{1/3}\,.
\end{equation}

\noindent Since the phenomena of strong lensing (e.g. multiple images, arcs,
$\ldots$) requires $\kappa_c>1$, equation (\ref{kclcdm}) indicates that,
for our illustrative choice of $(z_L,z_S)=(0.612,3)$
in a $\Lambda$CDM universe,
strong lensing requires lensing halos as massive
as $M\ga1.5\times10^{15}{\rm M_\odot}$. We can express this more generally by 
evaluating equations (\ref{sigmacrit}) and (\ref{sigma0}) at
different values of $z_L=z_{\rm coll}$ for a given source redshift $z_S$,
as shown in Figure~12. The quantity $\kappa_c(M/10^{15}{\rm M_\odot})^{-1/3}$
has a maximum value given by
\begin{equation}
\left[\kappa_cM_{15}^{-1/3}\right]_{\max}(z_S=3)=1.1686\,,
\end{equation}

\begin{figure}
\vspace{-10pt}
\hspace{-10pt}
\includegraphics[width=90mm]{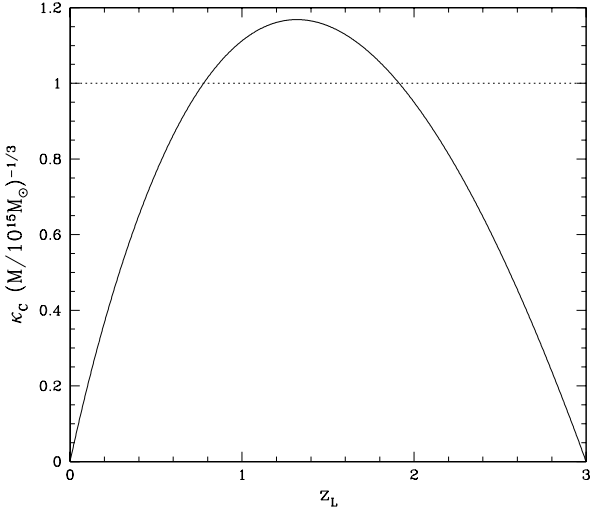}
\vspace{-45pt}
 \caption{The central convergence $\kappa_c$
[divided by the factor $(M/10^{15}{\rm M_\odot})^{1/3}$ which contains
the dependence on halo mass $M$] of TIS halos in the $\Lambda$CDM
universe versus lens redshift $z_L$, for the illustrative case in which
source is located at redshift $z_S=3$. The horizontal dotted line
indicates the minimum value ($\kappa_c=1$) required for strong lensing
by cluster-mass halos of $10^{15}{\rm M_\odot}$. (Such halos can only
produce strong lensing for $\kappa_c>1$, i.e. $0.777<z_L<1.914$.)}
\end{figure}

\noindent for the $\Lambda$CDM universe (which occurs at $z_L=1.320$),
where $M_{15}\equiv M/10^{15}{\rm M_\odot}$. The condition $\kappa_c>1$,
therefore, requires \begin{equation}
M_{15}\geq0.6266\,.
\end{equation}

To see how likely this is, let us focus on our illustrative example
of $(z_L,z_S)=(0.612,3)$.
For the cosmological model we consider, a 1-$\sigma$ density
fluctuation collapsing at redshift $z_{\rm coll}=0.612$ has a mass of about
$1.5\times10^{12}{\rm M_\odot}$ (the precise value depends on the
details of the power spectrum and its normalization).
Such ``typical'' objects will not be capable of
producing multiple images of a source at redshift $z_S=3$, since the resulting
value $\kappa_c=0.100$ is smaller than unity. 
This simply indicates that
multiple images are not produced by typical objects,
which is certainly consistent with the fact that fewer than 100 multiple-image
systems have been observed. Increasing $\kappa_c$ above
unity would require an object of mass
$M\approx1.5\times10^{15}{\rm M_\odot}$, a thousand times
the mass of a typical object at that redshift. Clusters of this mass
are rare but do exist. We can make a simple estimate of how atypical
such a massive object is. Over most of the mass range of cosmological
interest (from small
galaxies to clusters of galaxies) the CDM power spectrum can
be roughly approximated
by a power law $P(k)\propto k^n$, where $k$ is the wavenumber
and $n\approx-2$. The rms density 
fluctuation $\delta_{\rm rms}$ is then given by 
$\delta_{\rm rms}\approx k^{3/2}P^{1/2}(k)\propto k^{1/2}$.
At a given redshift, different values of
the wavenumber $k$ correspond to different mass scales 
$M$ according to $M\propto k^{-3}$. The relation between rms
density fluctuation and mass scale at fixed epoch is therefore approximated by
\begin{equation}
\delta_{\rm rms}\propto M^{-1/6}\,.
\end{equation}

\noindent Increasing the mass by a factor of 1000 therefore reduces
$\delta_{\rm rms}$ 
by a factor of $1000^{1/6}\approx3$. Because of the reduction
in $\delta_{\rm rms}$, 
a 1-$\sigma$ fluctuation ($\delta=\delta_{\rm rms}$) at this higher mass
will no longer collapse by this redshift $z_L=0.612$,
but a 3-$\sigma$ fluctuation ($\delta=3\delta_{\rm rms}$)
will. Such fluctuations are rare, but not
vanishingly rare. In Gaussian statistics, the probability that a randomly
located point in space is inside a 3-$\sigma$ density fluctuation
(i.e. $\delta\geq3\delta_{\rm rms}$) is 1/384.
Of course, whether any halo produced by such a fluctuation will
actually produce multiple images
will depend on the location of the sources. We carried
this simple calculation for
illustration purpose only. In a future paper, we will present a detailed
calculation of the expected frequency of multiple image systems for
comparison with the statistics of observed lensing.

In the TIS model, there is a relationship between the parameters $\rho_0$
and $r_0$, and the total mass $M$ of the halo, given by
\begin{equation}
M=4\pi\rho_0r_0^3\tilde M_t\,,
\end{equation}

\noindent where $\tilde M_t=61.485$ (Paper I). Using this relation, we can
directly estimate the separations between multiple images. The Einstein
radius of a lens of mass $M$
is defined as the angular radius $\theta_E$ of an Einstein
ring produced by a Schwarzschild lens\footnote{A point mass.} of
the same mass,
\begin{equation}
\label{thetae}
\theta_E=\left({4GM\over c^2}{D_{LS}\over D_LD_S}\right)^{1/2}
\end{equation}

\noindent (SEF, eq.~[2.6a]). This Einstein radius is often used to estimate
the characteristic scale of image features caused by strong lensing
(e.g. ring radius, radial location of arcs, image separations) and to
estimate the size of the region within which the mass which is
responsible for that strong lensing must be concentrated.
Since lensing halos are not actually point masses, however,
the angular radius $\theta_{\rm ring}$ of the 
actual Einstein ring which results if the source is located along the line of sight through the lens center will usually
differ from the Einstein radius $\theta_E$, assuming that the lens mass
distribution is actually capable of producing a ring.
The angular radius $\theta_{\rm ring}$ of the actual Einstein ring
produced by a TIS is given by
\begin{equation}
\label{thetaring}
\theta_{\rm ring}={r_0x_t\over D_L}\,,
\end{equation}

\begin{figure}
\vspace{-48pt}
\hspace{-20pt}
\includegraphics[width=100mm]{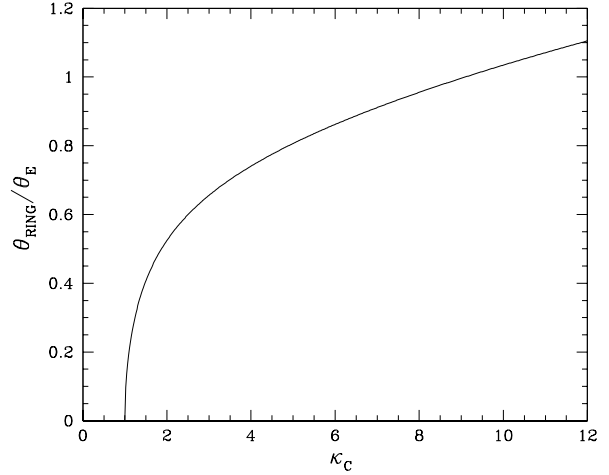}
\vspace{-45pt}
 \caption{Ratio of the radius $\theta_{\rm ring}$ of an Einstein ring
produced by a TIS halo to the radius $\theta_{\rm E}$ of an Einstein ring
produced by a Schwarzschild lens of the same mass (the Einstein radius),
versus $\kappa_c$.}
\end{figure}

\noindent where $x_t$ is the solution of equation~(\ref{tancrit}).
We take the ratio of equations~(\ref{thetae}) and~(\ref{thetaring}),
and use equation~(\ref{sigmacrit}) to eliminate the angular diameter
distances. We get, after some algebra,
\begin{eqnarray}
\label{thethe}
{\theta_{\rm ring}\over\theta_E}\!\!\!\!&=&\!\!\!\!
\left({\pi\over4}\right)^{1/2}
\left({A\over a}-{B\over b}\right)^{1/2}\tilde M_t^{-1/2}
\kappa_c^{-1/2}x_t \nonumber \\
&&\qquad\qquad\qquad=0.1565\kappa_c^{-1/2}x_t\,.
\end{eqnarray}

\noindent In Figure~13, we plot $\theta_{\rm ring}/\theta_E$ as
a function of $\kappa_c$.
A ring can be produced only if $\kappa_c>1$. The ratio of
the actual ring radius to the Einstein radius monotonically
increases with $\kappa_c$, and exceeds unity for $\kappa_c>9.096$.
Hence, the ring produced by a TIS can be either larger or smaller than
the ring produced by a Schwarzschild lens of the same mass, depending on
the value of $\kappa_c$. In the limit of large $\kappa_c$,
we have $\theta_{\rm ring}/\theta_E\approx0.2563\kappa_c^{1/2}$
according to equations~(\ref{xt2}) and~(\ref{thethe}).
We saw in \S4.1 that in cases of multiple images, the separation
$\Delta x$ between the outermost images is always of order $2x_t$.
Hence, $\theta_{\rm ring}$ can be reinterpreted as being one half the 
separation between images.

\section{SUMMARY AND CONCLUSION}

We have derived the lensing properties of cosmological halos
described by the Truncated Isothermal Sphere model. The solutions depend
on the background cosmological model through the critical surface
density $\sigma_{\rm crit}$, which is a function of the cosmological
parameters and the source and lens redshifts, and the TIS parameters
$\rho_0$ and $r_0$, which are functions of the mass and collapse redshift
of the halo, and the cosmological parameters. By expressing the surface
density of the halo in units of $\sigma_{\rm crit}$ and the distances
in units of $r_0$, all explicit dependences on the cosmological
model disappear, and the solutions are entirely expressible in terms of two
dimensionless parameters, the central convergence $\kappa_c$ and
the scaled position $y$ of the source. We have computed
solutions, and we provide either analytical expressions or numerical fits,
for the critical curves and caustics, the image separations, the
magnification and brightness ratios, the shear, and the time delay.
Lensing of a point source
by a TIS produces either one or three images, depending on
whether the source is located outside or inside the radial caustic. When
three images are produced, the central one is usually very faint, being
highly demagnified. Degenerate image configurations occur when an
extended source overlaps a caustic. Two images are produced when the
source overlaps the radial caustic, while an Einstein ring with a
central spot is produced when the source overlaps the tangential caustic,
which is a single point located at $y_t=0$.
These degenerate cases correspond to maxima of the total
magnification, which diverges as the source size goes to zero. When
three images are produced, the angular separation between the
two outermost images depends strongly on $\kappa_c$, but only
weakly on the source location.

The lens properties are often qualitatively different at small and 
large $\kappa_c$. For instance, at small $\kappa_c$, $\kappa_c\ga1$,
the image separation
$\Delta x$ decreases as source position $y$ increases
(i.e. as the projected separation between the source position
and the lens center increases)
(Fig.~7), the brightness ratio
$\mu_1/\mu_3$ increases with $y$ (Fig.~8), the shear $\gamma_3$ of the
outermost image decreases with $y$,
and the time delay $\tau_{23}$
decreases with $y$ for $y\la y_r$ (Fig.~10), while these quantities
behave differently for large $\kappa_c$. This is easily understood. Multiple
imaging can only occur when the central surface density $\sigma(0)$
exceeds the critical density $\sigma_{\rm crit}$ (or equivalently 
$\kappa_c>1$). Since $\sigma(x)$ is a decreasing function of $x$, there
is a natural scale in the system: the position $x_{\rm crit}$ on the
lens plane where $\sigma(x_{\rm crit})=\sigma_{\rm crit}$. 
If the lens profile was
scale-free, as in the cases of a Schwarzschild lens
or a singular isothermal sphere, $x_{\rm crit}$ would be the only length
scale in the problem, and the properties of the lens would be self-similar.
But the TIS has a characteristic length scale, the core radius $r_0$, and the
existence of this second length scale prevents the solutions from
being self-similar. 

This paper focused on the intrinsic properties of individual lenses
described by the TIS model. It provides all the necessary formula
one needs to study gravitational lensing by TIS halos in specific
cosmological models. We will present such studies in forthcoming papers.
As an illustration here, we applied the TIS model to the currently-favored
$\Lambda$CDM universe, to calculate the central convergence $\kappa_c$
expected for TIS halos of different masses and collapse epochs.
We found that high-redshift sources (e.g. $z_S\approx3$) will be
strongly lensed by TIS halos (i.e. $\kappa_c>1$) only for cluster-mass halos. 
We also calculated the characteristic angular scale of image features
produced by strong lensing by TIS halos relative to the Einstein radius
$\theta_E$ of a lens with the same total mass, for comparison
with the results for other lensing halo mass profiles and with observed lensing
systems.

\section*{Acknowledgments}

This work was supported by NASA ATP Grants NAG5-10825 and NAG5-10826,
and Texas Advanced Research Program
Grant 3658-0624-1999.

\bsp

\label{lastpage}

\end{document}